\documentclass[a4paper,11pt]{article}
\pdfoutput=1
\usepackage{jheppub}
\usepackage{amsthm,amsmath,amssymb,amscd}
\usepackage{amsfonts,bm,latexsym}
\usepackage{xcolor}
\usepackage{orcidlink}

\def\be{\begin{equation}}
\def\ee{\end{equation}}
\def\bea{\begin{eqnarray}}
\def\eea{\end{eqnarray}}
\def\nn{\nonumber}
\def\p{\partial}
\def\ta{\theta}
\def\cta2{\cos^2\theta}
\def\sta2{\sin^2\theta}
\def\bA{\mathbf{A}}
\def\cA{\mathcal{A}}
\def\ctA{\mathcal{\tA}}
\def\cF{\mathcal{F}}
\def\cN{\mathcal{N}}
\def\cP{\mathcal{P}}
\def\tA{\widetilde{A}}
\def\tcA{\widetilde{\cA}}
\def\tF{\widetilde{F}}
\def\tcF{\widetilde{\cF}}
\def\ctF{\mathcal{\tF}}
\def\Star1{\,\star\, 1\!\!1}

\title{General non-extremal rotating charged AdS black hole solutions with
four independent charge parameters in $D=4$, STU gauged supergravity}
\author[a,b]{Di Wu\orcidlink{0000-0002-2509-6729}}
\author[a]{and Shuang-Qing Wu\orcidlink{0000-0001-7936-7195}}
\affiliation[a]{School of Physics and Astronomy, China West Normal University,\\
1 Shida Road, Nanchong, Sichuan 637002, People's Republic of China}
\affiliation[b]{Department of Physics and Astronomy, University of Waterloo,\\
200 University Ave W, Waterloo, ON N2L 3G1, Canada}
\emailAdd{wdcwnu@163.com}
\emailAdd{sqwu@cwnu.edu.cn}
\arxivnumber{2611.13151}

\abstract{
We present an exact solution for generic non-extremal rotating, charged black holes with a negative
cosmological constant in four-dimensional $\mathcal{N}=2, U(1)^4$ gauged supergravity. This general
non-extremal solution contains seven independent parameters related to the mass, rotation, two electric
and two magnetic charges, as well as a nonzero cosmological constant. It generalizes the four-dimensional
Cveti\v{c}-Youm black hole solution to the gauged supergravity case. We anticipate that this is the only
regular black hole solution in four-dimensional gauged STU model without any Misner string singularity
and/or C-metric-like conical defect on the symmetry axes. We also investigate its thermodynamic property
and give the Killing-St\"ackel tensor responsible for the separability of Hamilton-Jacobi and massless
Klein-Gordon equations. Various special cases related to the known solution are discussed, and its BPS
limit is also analyzed. Our solution provides a new background to test the AdS$_4$/CFT$_3$ correspondence
and to study the microscopic entropy via extremization of the BPS superconformal index.
}

\begin{document}
\maketitle
\flushbottom

\section{Introduction}\label{sec:introduction}

Rotating charged black holes in anti-de Sitter (AdS) spacetime play an important role in the AdS/CFT
correspondence, holography, and supersymmetric localization \cite{LRR23-2}. On the field theory side
of this correspondence, black hole thermodynamics is reflected in statistical ensembles characterized
by angular momentum, chemical potentials, and magnetic fluxes. In particular, microscopic state counting
through supersymmetric localization and related field theory indices has reproduced the Bekenstein-Hawking
entropy of supersymmetric AdS$_4$ black holes, including static electrically and magnetically charged,
rotating electrically charged, and rotating and accelerating configurations \cite{JHEP0516054,PLB771-462,
JHEP0320081,PRL133-031603}. These connections motivate the construction of exact rotating AdS black holes
carrying independent electric and magnetic charges. However, constructing general non-extremal charged
rotating AdS black holes with multiple independent charges and/or rotation parameters remains very
difficult since no general solution-generating technique is available. An important development was the
extension of the familiar Kerr-Schild metric ansatz to non-extremal charged and/or rotating black
holes in supergravity theories \cite{PRD83-121502,PLB705-383}, leading to general rotating charged
solutions in $D=5$ and $D=7$ gauged supergravities \cite{PLB707-286, JHEP0923143}. In $D=4$, such an
ansatz can be applied solely to purely electrically or purely magnetically charged sectors, while
solutions carrying both types of charges might call for a treatment incorporating generalized
electromagnetic duality in the STU model.

In the four-dimensional gauged STU model, the known rotating charged solutions include pairwise
equal-charge cases \cite{NPB717-246,PRD89-065003} (see also \cite{FN}) and single-charge
\cite{PRD83-121502,CQG28-032001} and two-charge \cite{CQG28-175004} solutions obtained through
consistent truncations. In the ungauged theory, the pairwise equal-charge case includes the Kerr-Sen
black hole of Einstein-Maxwell-dilaton-axion (EMDA) theory \cite{PRL69-1006}, its partial extension
\cite{PRL74-1276}, and the Kerr--Newman black hole with equal electric and magnetic charges; see
Ref. \cite{PRD103-044014} for a brief review of the EMDA solutions. These known solutions are
often expressed in complicated forms involving charging parameters \cite{NPB717-246,PRD89-065003}.
More general rotating black hole solutions carrying four independent electric and four independent
magnetic charge parameters have been constructed in the ungauged STU model \cite{PR316-1,CQG31-022001,
PRD90-025029}. Introducing a negative cosmological constant changes the asymptotic geometry,
thermodynamic properties, and supersymmetric structure of rotating black holes with independent
electric and magnetic charges.

In this work, we present an exact non-extremal rotating charged AdS black hole solution in the
four-dimensional $U(1)^4$ gauged supergravity, carrying two independent electric and two independent
magnetic charge parameters. It extends the Cveti\v{c}-Youm solution \cite{PRD54-2612} to a nonzero
cosmological constant. We give the metric, gauge potentials, and scalar fields in a unified form
and show how previously known rotating black holes arise as special cases. We then investigate its
thermodynamic properties and the BPS limit, connecting the non-extremal solution to its extremal
case. The resulting family provides an exact setting for studying the combined effects of rotation
and independent electromagnetic charges in AdS black hole physics, with potential applications to
charged thermal and supersymmetric states in holography and localization.

\section{The $D=4$, $\mathcal{N}=2$ gauged STU model}\label{sec:theory}

The maximal four-dimensional $\cN=8, SO(8)$ gauged supergravity can be consistently truncated to
$U(1)^4$ gauged supergravity, which is $\cN =2$ gauged supergravity coupled to three Abelian vector
multiplets \cite{JGP23-111,NPB558-96,NPB554-237}. The bosonic fields of this truncation are the
graviton $g_{\mu\nu}$, four U(1) gauge fields $A_I$, three dilatons $\varphi_i$ and three axions
$\chi_i$, with $I=i, 4$ and $i=1, 2, 3$.

The four-dimensional Lagrangian for the bosonic sector of the $\cN=2$, STU supergravity coupled to
three vector multiplets can be written as
\be\begin{aligned}
\mathcal{L} &= R\Star1 +g^2\sum_{i=1}^3\big[e^{-\varphi_i} +e^{\varphi_i}
\big(1 +\chi_i^2\big)\big]\Star1
 -\frac{1}{2}\sum_{i=1}^3\big({^\star\!}d\varphi_i \wedge d\varphi_i
 +e^{2\varphi_i}\,{^\star\!}d\chi_i \wedge d\chi_i\big) \\
& -\frac{1}{2}\big(X_1{^\star\!}F_1\wedge F_1 +X_2{^\star\!}F_2 \wedge F_2
 +X_3{^\star\!}\cF_1 \wedge \cF_1
  +X_4{^\star\!}\cF_2 \wedge \cF_2\big) \\
& -\chi_2(F_1 \wedge \cF_1 +F_2\wedge \cF_2) \, ,
\end{aligned}\ee
where
\bea
&& X_1 = e^{\varphi_1 -\varphi_2 -\varphi_3} \, , \quad
X_2 = e^{\varphi_1 -\varphi_2 +\varphi_3} \, , \nn \\
&& X_3 = e^{-\varphi_1 -\varphi_2 +\varphi_3} \, , \quad
X_4 = e^{-\varphi_1 -\varphi_2 -\varphi_3} \, . \nn
\eea
In these expressions, we have made the following changes of notations relative to the convention
of Ref. \cite{NPB717-246} in order to cast the scalar fields into a unified fashion: $(\varphi_1,
\varphi_2) \to (\varphi_2, \varphi_1), \quad (\chi_1, \chi_2) \to (\chi_2, -\chi_1)$.

The four Maxwell field strengths and their corresponding dual ones are defined in terms of vector
potentials as follows:
\bea
F_1 &=& dA_1 +\chi_1\,d\cA_2 \, , \nn \\
F_2 &=& dA_2 -\chi_3\,dA_1 -\chi_1\,d\cA_1 -\chi_1\chi_3\,d\cA_2 \, , \nn \\
\cF_1 &=& d\cA_1 +\chi_3\,d\cA_2 \, ,\quad \cF_2 = d\cA_2 \, , \nn \\
\tF_1 &=& d\tA_1 +\chi_3\,d\tA_2 \, , \quad \tF_2 = d\tA_2 \, , \nn \\
\ctF_1 &=& d\ctA_1 +\chi_1\,d\tA_2 \, , \nn \\
\ctF_2 &=& d\ctA_2 -\chi_3\,d\ctA_1 -\chi_1\,d\tA_1 -\chi_1\chi_3\,d\tA_2 \, , \nn
\eea
so that the generalized electro-magnetic duality can be concisely written as
\be\begin{aligned}
& X_1{^\star\!}F_1 +\chi_2\cF_1 = \tF_1 \, , \quad  X_2{^\star\!}F_2 +\chi_2\cF_2 = \tF_2 \, , \\
& X_3{^\star\!}\cF_1 +\chi_2F_1 = \tcF_1 \, , \quad X_4{^\star\!}\cF_2 +\chi_2F_2 = \tcF_2 \, .
\label{gemd}
\end{aligned}\ee

\section{The general non-extremal rotating solution}\label{sec:solution}

In Ref. \cite{PRD54-2612}, Cveti\v{c} and Youm only gave the metric of a four-charged rotating
black hole solution with each two independent electric and magnetic parameters. Later, the complete
solution was regenerated by Chong, \textit{et al} in Ref. \cite{NPB717-246} from the seed Kerr
solution and presented the explicit expressions of four gauge potentials. In 2014, its AdS extension
was constructed by adding a nonzero cosmological constant and reported twice by one of us but still
unpublished.

This general non-extremal solution comprises the metric, four gauge potentials and their magnetic
duals, and three dilaton--axion pairs in a unified parametrization. It is specified by seven
independent solution parameters: the mass parameter $m$, rotation parameter $a$, and four independent
charge parameters $\delta_I$ as well as the gauge-coupling constant $g$, so it describes a family
of nonextremal rotating anti--de Sitter black holes with two different electric charges and two
different magnetic charges in the four-dimensional $U(1)^4$ gauged supergravity.

Its line element can be explicitly written in a compact form as follows:
\bea
ds^2 &=& -\frac{H\Delta_\ta}{\sqrt{W}\Xi}\Big[dt
 +\frac{\omega(r)}{H} a\sta2\, d\phi\Big]^2
 +\sqrt{W}\bigg[\frac{dr^2}{\Delta(r)} +\frac{d\ta^2}{\Delta_\ta}
 +\frac{\Delta(r)}{H\Xi}\sta2\, d\phi^2\bigg] \, . \label{tdr}
\eea
This metric is initially obtained via the dimensional reduction form \cite{JHEP0114127} along the
timelike direction (\ref{tdr}). It can not be cast into Wu's ansatz because $W$ (given below) can
not be put into multiply form of several simpler factors.

Three dilatons and three axions can be written in a unified manner:
\be\begin{aligned}
e^{\varphi_i} &= \frac{r_1r_2r_3/r_i +a^2\cta2}{\sqrt{W}} \, , \\
\chi_i &= \frac{2mc_is_ia\cos\ta}{r_1r_2r_3/r_i +a^2\cta2}\Big(\frac{c_1c_2c_3s_4}{c_i^2}
 -\frac{s_1s_2s_3c_4}{s_i^2}\Big) \\
&= \frac{2mc_is_ia\cos\ta}{r_1r_2r_3/r_i +a^2\cta2}\Big(\frac{s_4\alpha_i}{c_4}
 -\frac{c_4\beta_i}{s_4}\Big) \, ,
\end{aligned}\ee
with the lowercase index running over $i = 1, 2, 3$.

We also use the index $I=i, 4$ to label the four gauge potentials and their dual counterparts which
are regrouped as the `electric-type' potentials $\bA_I = (\tA_1, A_2, \tcA_1, \cA_2)$, and the
`magnetic-type' potentials $\widetilde{\bA}_I = (A_1, -\tA_2, \cA_1, -\tcA_2)$:
\bea
\bA_I &=& \frac{2mc_Is_I}{W\sqrt{\Xi}}\Big\{\big(r_1r_2r_3r_4/r_I
 +ra^2\cta2\big)\big[\Delta_\ta\, dt -(\alpha_I -\beta_I)a\sta2\, d\phi\big] \nn \\
&&+2ma^2\cta2\big(w_I\Delta_\ta\, dt -\beta_I a\sta2\, d\phi\big)\Big\} \, , \\
\widetilde{\bA}_I &=& \frac{2mc_Is_Ia\cos\ta}{W\sqrt{\Xi}}\Big\{\big(rr_I
 +a^2\cta2\big)\big[(\alpha_I -\beta_I)\Delta_\ta\, dt
-a\sta2\, d\phi\big] \nn \\
&&+2mr_I\big(\beta_I\Delta_\ta\, dt -w_I a\sta2\, d\phi\big)\Big\}
-\frac{2mc_Is_I\cos\ta}{\sqrt{\Xi}}\big[d\phi -(\alpha_I -\beta_I)g^2a\, dt\big] \, .
\eea
They are related via the generalized electro-magnetic duality in Eq. (\ref{gemd}).

Various auxiliary functions and notations entering the solution are given below:
\bea
 H &=& \big(r^2 +a^2\cta2 -2mr\big)\big[1 +g^2K(r)\big] +g^2\omega(r)^2 \nn \\
 &=& \Delta(r) -a^2\sta2\big[1 +g^2K(r)\big] \, , \nn \\
\Delta(r) &=& \big(r^2 +a^2 -2mr\big)\big[1 +g^2K(r)\big] +g^2\omega(r)^2 \, , \nn
\eea
and in terms of $r_I = r +2ms_I^2$, $c_I = \cosh\delta_I$, and $s_I = \sinh\delta_I$,
the remaining expressions read
\bea
W &=& \big(r_1r_2 +a^2\cta2\big)\big(r_3r_4 +a^2\cta2\big)
-4m^2(c_1c_2s_3s_4 -s_1s_2c_3c_4)^2a^2\cta2 \, , \nn \\
K(r) &=& r_1r_2 +r_3r_4 -r^2 +2mr
-4m^2(c_1c_2s_3s_4 -s_1s_2c_3c_4)^2 \, , \nn \\
\omega(r) &=& 2mc_1c_2c_3c_4r -2ms_1s_2s_3s_4(r -2m) \, ,
\quad \alpha_I = c_1c_2c_3c_4c_I^{-2} \, , \quad \beta_I = s_1s_2s_3s_4s_I^{-2} \, , \nn \\
w_I &=& 1 +s_1^2 +s_2^2 +s_3^2 +s_4^2 -s_I^2
+(s_1s_2s_3s_4 -c_1c_2c_3c_4)(\alpha_I -\beta_I) \, . \nn
\eea
Here and hereafter, we also denote $\Delta_\ta = 1 -g^2a^2\cta2$, $\Xi = 1 -g^2a^2$, and
$\Sigma = r^2 +a^2\cta2$ as well as $r_i = r +2ms_i^2$ throughout this article.

\subsection{Equivalent forms of the metric}\label{sec:equiv}

The line element (\ref{tdr}) also admits the equivalent representation
\bea
ds^2 &=& -\frac{r^2+a^2\cta2 -2mr}{\sqrt{W}\Xi} \Big[\Delta_\ta dt
 +\frac{\omega(r)a\sta2}{r^2+a^2\cta2 -2mr} d\phi\Big]^2 +\frac{d\ta^2}{\Delta_\ta} \nn \\
&&\quad~ +\sqrt{W}\bigg[-\frac{g^2\Delta_\ta}{\Xi} dt^2 +\frac{dr^2}{\Delta(r)}
+\frac{\big(r^2+a^2-2mr\big)\sta2}{\big(r^2+a^2\cta2 -2mr\big)\Xi} d\phi^2\bigg] \, ,
\eea
which is more easily related to its ungauged counterpart. Alternatively, it can be also rewritten as
\bea
ds^2 &=& -\frac{\Delta(r)\Delta_\ta}{\sqrt{W}\Xi} dt^2 +\sqrt{W}\Big(\frac{dr^2}{\Delta(r)}
 +\frac{d\ta^2}{\Delta_\ta}\Big) +\frac{\Delta_\ta\sta2\big(a\big[1 +g^2K(r)\big]dt
 -\omega(r) d\phi\big)^2}{\sqrt{W}[1 +g^2K(r)]\Xi} \nn \\
&& +\frac{\Delta(r)[1 +g^2K(r) -\Delta_\ta]}{g^2\sqrt{W}\big[1 +g^2K(r)\big]\Xi}\sta2\,d\phi^2 \, ,
\eea
whose inverse metric has a similar compact structure, and its components are elegantly given by
\bea
\sqrt{W}\, \p_s^2 &=& \Delta(r)\p_r^2 +\Delta_\ta\p_{\ta}^2 +\frac{\Xi}{\sta2}\p_{\phi}^2
+\frac{\Xi}{g^2}\Big[\frac{-1}{\Delta_\ta} +\frac{1}{1 +g^2K(r)}\Big]\p_t^2 \nn \\
&& -\Xi\frac{\big(\omega(r)\p_t +a\big[1 +g^2K(r)\big]\p_{\phi}\big)^2}{\Delta(r)
 \big[1 +g^2K(r)\big]} \, , \qquad
\eea
with which the Hamilton-Jacobi and massless Klein-Gordon equations can be easily written down.

\section{Special cases}\label{sec:special}

We first discuss several limits of the general solution, followed by the pairwise-equal and two-charge sectors.Each provides a direct consistency check of our construction.
First, the nonrotating case ($a = 0$) directly reproduces that was discussed in Ref. \cite{NPB554-237}.
Second, in the \emph{asymptotic flat case} ($g = 0$), the Cveti\v{c}-Youm black hole \cite{NPB717-246,
PRD54-2612} is recovered by directly setting $g = 0$.

Next, we turn to the uncharged case ($\delta_I = 0$). With all charge parameters are set to zero,
the solution becomes the Kerr-AdS black hole whose metric expressed within the Kerr-Schild form is
\bea
ds^2 &=& -\frac{1 +g^2r^2}{\Xi}\Delta_\ta\, dt^2 +\frac{r^2+a^2}{\Xi}\sta2\, d\phi^2
 +\frac{\Sigma}{\bar{\Delta}_r} dr^2 \nn \\
&& +\frac{\Sigma}{\Delta_\ta} d\ta^2 +\frac{2mr}{\Sigma\Xi}
 \big(\Delta_\ta\, dt -a\sta2\, d\phi\big)^2 \, ,
\eea
where $\bar{\Delta}_r = (1 +g^2r^2)(r^2+a^2) -2\Xi\, mr$. One can recognize that it is the ordinary
one whose mass parameter is replaced by $m\to \mathcal{M} = m\Xi$.

\subsection{Pairwise equal-charge case}\label{sec:equalcharges}

For $\delta_3=\delta_1$ and $\delta_4=\delta_2$,
one has $\varphi_1 = \varphi_3 = 0$, $\chi_1 = \chi_3 = 0$, and the metric reduces to
\be
ds^2 = -\frac{\hat{\Delta}_r}{\hat{W}}X^2 +\hat{W}\Big(\frac{dr^2}{\hat{\Delta}_r}
 +\frac{d\theta^2}{\Delta_{\ta}}\Big) +\frac{\Delta_{\ta}\sta2}{\hat{W}}Y^2 \, ,
\ee
where $\hat{W} = r_1r_2 +a^2\cta2$ and
\be\begin{aligned}
\hat{\Delta}_r &= (r^2+a^2 -2mr)\Xi +g^2(r_1r_2 +a^2)^2 \, , \qquad
X = \frac{\Delta_{\ta}}{\Xi}dt -\frac{a\sta2}{\Xi}d\phi \, , \\
Y &= \frac{1 +g^2r_1r_2}{\Xi}a dt -\frac{r_1r_2 +a^2}{\Xi}d\phi \, . \nn
\end{aligned}\ee

The four Abelian vector potentials and their duals are
\be\begin{aligned}
& \cA_1 = A_1 = \frac{2\sqrt{\Xi}mc_1s_1\cos\ta}{\hat{W}}Y \, , \qquad \cA_2 = A_2 = \frac{2\sqrt{\Xi}mc_2s_2r_1}{\hat{W}}X \, , \\
& \tilde{\cA}_1 = \tilde{A}_1 = \frac{2\sqrt{\Xi}mc_1s_1r_2}{\hat{W}}X \, , \qquad
 \tilde{\cA}_2 = \tilde{A}_2 = -\frac{2\sqrt{\Xi}mc_2s_2\cos\ta}{\hat{W}}Y \, ,
\end{aligned}\ee
and the two surviving scalar fields are
\be\begin{aligned}
\varphi_2 = \ln\Big(\frac{r_1^2+a^2\cta2}{\hat{W}}\Big) \, , \qquad
\chi_2 = \frac{2m\big(s_2^2-s_1^2\big)a\cos\ta}{r_1^2+a^2\cta2} \, ,
\end{aligned}\ee
This solution is comparable with that was given in Ref. \cite{NPB717-246,hep-th/0504080}, whose
uncharged version is the usual Kerr-AdS solution with the mass parameter $m$. In order to
find their relationship, we first need to rescale the time and azimuthal coordinates and shift
the latter one so that the Killing parts have a similar structure in terms of $X$ and $Y$. Then
they coincide exactly with each other after shifting the radial coordinate by a constant followed
by properly redefining the mass and charge parameters. One will obtain another form of the
Kerr-Sen-AdS solution if $\delta_1 = 0$ and its magnetic dual by setting $\delta_2 = 0$. The
Kerr-Newman-AdS solution with equal electric and magnetic charge is reproduced in another form
if $\delta_2 = \delta_1 = \delta$, which is different from the most general Kerr-Newman-AdS
solution with unequal electric and magnetic charges \cite{CQG17-399}.

\subsection{Two-charge case}\label{sec:twocharges}

For $\delta_1 = \delta_3 = 0$, in this purely electric two-charge sector,
one has: $A_1 = \cA_1 = \tilde{A}_1 = \tilde{\cA}_1 = 0$, and $\chi_1 = \chi_3 = 0$. The metric
can be recast into the Wu's ansatz as follows:
\bea
ds^2 &=& \sqrt{H_2H_4}\bigg[-\frac{\big(1+g^2r^2\big)\Delta_{\ta}}{\Xi} dt^2
 +\Sigma\Big(\frac{dr^2}{\Delta_r}
 +\frac{d\theta^2}{\Delta_{\ta}}\Big) +\frac{\big(r^2+a^2\big)\sta2}{\Xi} d\phi^2 \nn \\
&& +\frac{2mrs_2^2\Xi}{H_2\Sigma\big(s_2^2-s_4^2\big)}K_2^2
 +\frac{2mrs_4^2\Xi}{H_4\Sigma\big(s_4^2-s_2^2\big)}K_4^2\bigg] \, . \label{Wuma} \qquad
\eea
where we denote
$$K_i = \frac{c_i\Delta_{\ta}}{\Xi}dt -\frac{c_2c_4a\sta2}{c_i\Xi} d\phi \, , \qquad
(i = 2, 4) \, .$$
The nonzero two vector potentials and their duals are
\bea
&& A_2 = \frac{2ms_2r\sqrt{\Xi}}{H_2\Sigma}K_2 \, , \quad
\cA_2 = \frac{2ms_4r\sqrt{\Xi}}{H_4\Sigma}K_4 \, , \nn \\
&& \tilde{A}_2 = -\frac{2ms_2\cos\ta}{H_4\Sigma\sqrt{\Xi}}\big[c_4\big(1 +g^2rr_4\big)a dt
 -c_2\big(rr_4 +a^2\big)d\phi\big] \, , \nn \\
&& \tilde{\cA}_2 = -\frac{2ms_4\cos\ta}{H_2\Sigma\sqrt{\Xi}}\big[c_2\big(1 +g^2rr_2\big)a dt
 -c_4\big(rr_2+a^2\big)d\phi\big] \, , \nn
\eea
and the scalar fields are
\bea
&& \varphi_1 = \varphi_3 = \frac{1}{2}\ln\Big(\frac{H_2}{H_4}\Big) \, , \quad
\varphi_2 = -\frac{1}{2}\ln(H_2H_4) \, ,  \quad
\chi_2 = \frac{2ms_2s_4a\cos\ta}{\Sigma} \, , \nn
\eea
where $H_i = 1 +2mrs_i^2/\Sigma$ and the radial function is
\bea
\Delta_r &=& (r^2+a^2 -2mr)\big[1 +g^2r^2 +2g^2mr\big(1 +s_2^2+s_4^2\big)\big] +4g^2m^2c_2^2c_4^2r^2 \, . \nn
\eea
The above expressions for our two-charge solution is much simpler than those obtained by Chow
in Ref. \cite{CQG28-175004}, whose metric can also be rewritten in the form (\ref{Wuma}) given
above. These two solutions are related via redefining the mass and charge parameters.

On the other hand, if one sets $\delta_2 = \delta_4 = 0$, then one gets the two purely
magnetic-charged rotating AdS black hole, which is related the above purely electric-charged
case via the electro-magnetic duality.

\section{Black hole thermodynamics}\label{sec:thermo}

The black hole solution presented in the above is regular simply because it contains no Misner string
and no conical defect on the symmetry axes. The azimuthal coordinate $\phi$ has the usual period of
$2\pi$, the radial coordinate locates in the range of $r\in (0, \infty)$. We shall focus on the case
that our solution has no naked singularity and contains no closed timelike curve by taking the suitable
intervals for solution parameters.

At the outer horizon $r_+$ which is the largest root of $\Delta(r_+) = 0$, the entropy, temperature,
and angular velocity take the form:
\be\begin{aligned}
S &= \frac{\pi\omega(r_+)}{\sqrt{\Xi[1 +g^2K(r_+)]}} \, , \\
T &= \frac{\sqrt{1 +g^2K(r_+)}}{4\pi\omega(r_+)\sqrt{\Xi}}\Delta^{\prime}(r_+) \, , \\
\Omega &= a\frac{1 +g^2K(r_+)}{\omega(r_+)} \, .
\end{aligned}\ee
The electric-static potentials read
\bea
\Phi_I &=& (\bA_I^{\mu}\zeta_{\mu})\big|_{r=r_+}
= \frac{2mc_Is_I}{\sqrt{\Xi}}\,\frac{r_1r_2r_3r_4/r_I
 +(r +2mw_I)a^2}{(r^2 +a^2 -2mr)K(r) +\omega(r)^2}\Big|_{r=r_+} \, . \qquad
\eea
while the magneto-static potentials vanish identically: $\widetilde{\Phi}_I = (\widetilde{\bA}_I^{\mu}
\zeta_{\mu})\big|_{r=r_+} = 0$, and $\zeta = \p_t +\Omega\p_{\phi}$ is the generator of the Killing
horizon.

Our solution is asymptotically AdS, whose boundary metric approaches to
\be
\frac{ds^2}{r^2} \mapsto -\frac{g^2\Delta_\ta}{\Xi} dt^2 +\frac{dr^2}{g^2r^4}
+\frac{d\ta^2}{\Delta_\ta} +\frac{\sta2}{\Xi} d\phi^2 \, .
\ee
Using the conformal completion method \cite{PRD73-104036}, the mass and angular momentum can be readily
calculated, while the four electromagnetic charges $\hat{Q}_I$ can be computed via the Gauss integral
and they are explicitly given by
\be\begin{aligned}
M &= \frac{m\big(2 +s_1^2 +s_2^2 +s_3^2 +s_4^2\big)}{2\Xi} \, , \\
J &= \frac{ma(c_1c_2c_3c_4 -s_1s_2s_3s_4)}{\Xi} \, , \\
\hat{Q}_I &= \frac{mc_Is_I}{2\sqrt{\Xi}} \, .
\end{aligned}\ee

The first law and Smarr relation in the extended thermodynamic phase space take the form
\bea
dM &=& T dS +\Omega\, dJ +V d\cP +\sum_{I=1}^4\Phi_I\, d\hat{Q}_I \, , \\
M &=& 2TS +2\Omega\, J -2V\cP +\sum_{I=1}^4\Phi_I\, \hat{Q}_I \, ,
\eea
in which the thermodynamic volume conjugate to the pressure  $\cP = 3g^2/(8\pi)$ is
\bea
V = \frac{\pi}{3\Xi}\bigg[\sum_{I=1}^4\frac{r_1r_2r_3r_4}{r_I}\Big|_{r=r_+} +4r_+a^2
 +4ma^2\big(1 +s_1^2 +s_2^2 +s_3^2 +s_4^2\big)\Big] \, .
\eea

\section{BPS limit}\label{sec:bps}

The BPS limit of the rotating charged solution (\ref{tdr}) is characterized by saturation of the
four-dimensional AdS Bogomol'nyi bound \cite{hep-th/0504080}: $M = gJ +\sum_{I=1}^4\hat{Q}_I$, for
positive charge parameters and positive rotation. Together with the degenerate-horizon condition,
this fixes the rotation and mass parameters as
\be
a \equiv b = \frac{1}{g\cosh D}\, , \quad  2m\equiv \mu = \frac{\sinh D}{g\sqrt{B\cosh D}} \, ,
\ee
where we have introduced the following two notations:
\be\begin{aligned}
D &= \delta_1 +\delta_2 +\delta_3 +\delta_4 \, , \\
B &= \sinh(\delta_1+\delta_2)\sinh(\delta_1+\delta_3)\sinh(\delta_1+\delta_4)
 \sinh(\delta_2+\delta_3)\sinh(\delta_2+\delta_4)\sinh(\delta_3+\delta_4)\, .
\end{aligned}\ee
in which the four charge parameters $\delta_I$ remain independent.

At the resulting extremal horizon, we can check that $\Omega = g$ and $\Phi_I = 1$. Also, the
Hawking temperature vanishes identically. The degenerate horizon is a double root of $\bar{\Delta}(r)$
and is located at
\be
r_0 = \frac{s_1s_2s_3s_4}{g\sqrt{B\cosh D}} \sum_{I=1}^4\frac{c_I}{s_I}
= \mu\frac{s_1s_2s_3s_4}{\sinh D} \sum_{I=1}^4\frac{c_I}{s_I} \, ,
\ee
obeying
\be
\bar{\Delta}(r_0) = 0 \, ,\qquad \bar{\Delta}^{\prime}(r_0) = 0 \, ,
\ee
where the radial function is
\bea
\bar{\Delta}(r) = (r -r_0)^2\Big(g^2r^2 +g^2v_1r +\frac{v_0}{r_0^2}\Big)\, ,
\label{bDr}
\eea
in which two coefficients are explicitly given by
\be\begin{aligned}
v_1 &= 2r_0 +\mu\big(s_1^2 +s_2^2 +s_3^2 +s_4^2\big) \, , \\
v_0 &= b^2 +g^2\mu^2b^2\big[s_1^2s_2^2 +s_3^2s_4^2 -(c_1c_2s_3s_4
 -s_1s_2c_3c_4)^2\big] +{g^2\mu^4}s_1^2s_2^2s_3^2s_4^2\, .
\end{aligned}\ee

The BPS geometry and conserved quantities are described in the next subsection.

\subsection{Geometry and thermodynamics of the BPS solution}\label{sec:bps-properties}

The BPS metric takes the same compact form as the general nonextremal solution (\ref{tdr}):
\bea
d\bar{s}^2 &=& -\frac{\bar{H}\bar{\Delta}_{\ta}}{\sqrt{\bar{W}}\bar{\Xi}}
 \Big[dt +\frac{\bar{\omega}(r)}{\bar{H}}b\sta2\,d\phi\Big]^2
+\sqrt{\bar{W}}\bigg[\frac{dr^2}{\bar{\Delta}(r)} +\frac{d\theta^2}{\bar{\Delta}_{\ta}}
 +\frac{\bar{\Delta}(r)\sta2}{\bar{H}\bar{\Xi}}\,d\phi^2\bigg] \, ,  \qquad
\label{bps-metric}
\eea
in which $\bar{\Delta}(r)$ is given by Eq. (\ref{bDr}) and $\bar{\Xi} = 1 -g^2b^2$. We have also
used $r_I = r +\mu{} s_I^2$ to write the following functions that characterize the solution:
\bea
\bar{W} &=& (r_1r_2 +b^2\cta2)(r_3r_4 +b^2\cta2)
 -\mu^2(c_1c_2s_3s_4 -s_1s_2c_3c_4)^2b^2\cta2\, , \nn \\
\bar{H} &=& \bar{\Delta}(r) -b^2\sta2\big[1 +g^2\bar{K}(r)\big] \, , \qquad
\bar{\Delta}_{\ta} = 1 -g^2b^2\cta2\, , \nn \\
\bar{K}(r) &=& r_1r_2 +r_3r_4 -r^2 +\mu{}r
-\mu^2(c_1c_2s_3s_4 -s_1s_2c_3c_4)^2 \, , \nn\\
\bar{\omega}(r) &=& \mu(c_1c_2c_3c_4-s_1s_2s_3s_4)r +\mu^2s_1s_2s_3s_4 \, . \nn
\eea

The three dilatons and three axions are kept in a unified form as
\be\begin{aligned}
e^{\bar{\varphi}_i} &= \frac{r_1r_2r_3/r_i +b^2\cta2}{\sqrt{\bar{W}}}\, ,\\
\bar{\chi}_i &= \frac{\mu{}c_is_i b\cos\ta}{r_1r_2r_3/r_i +b^2\cta2}
 \Big(\frac{s_4\alpha_i}{c_4} -\frac{c_4\beta_i}{s_4}\Big)\, .
\label{bps-scalars}
\end{aligned}\ee
In particular, the axions are nonzero in general.

The four `electric-type' potentials and their `magnetic-type' duals are
\bea
\bar{\bA}_I &=& \frac{\mu{}c_Is_I}{\bar{W}\sqrt{\bar{\Xi}}}
\Big\{\big(r_1r_2r_3r_4/r_I +rb^2\cta2\big)
 \big[\bar{\Delta}_{\ta}dt -(\alpha_I -\beta_I)b\sta2\,d\phi\big] \nn \\
&&\quad +\mu{}b^2\cta2\big[w_I\bar{\Delta}_{\ta}dt -\beta_Ib\sta2\,d\phi\big]\Big\} \, ,
\label{bps-electric} \\
\widetilde{\bar{\bA}}_I &=& \frac{\mu{}c_Is_Ib\cos\ta}{\bar{W}\sqrt{\bar{\Xi}}}
 \Big\{(rr_I +b^2\cta2)\big[(\alpha_I -\beta_I)\bar{\Delta}_{\ta}dt
 -b\sta2\,d\phi\big] \nn \\
&&+\mu{}r_I\big[\beta_I\bar{\Delta}_{\ta}dt
 -w_Ib\sta2\,d\phi\big]\Big\}
-\frac{\mu{}c_Is_I\cos\ta}{\sqrt{\bar{\Xi}}}
\big[d\phi -(\alpha_I -\beta_I)g^2b\,dt\big] \, .
\label{bps-magnetic}
\eea
Here $\alpha_I$, $\beta_I$, and $w_I$ retain their definitions in section~\ref{sec:solution}. Equations
(\ref{bps-metric}) and (\ref{bps-scalars})--(\ref{bps-magnetic}) specify the complete BPS field
configuration.

We now examine the geometry and conserved quantities of this BPS solution. The Killing vector
$k = \p_t +g\p_\phi$ has norm
\bea
k^2 &=& -\frac{(r -r_0)^2}{\sqrt{\bar{W}}} \Big\{\bar\Delta_{\ta}\sta2 
 +\Big(g^2r^2 +g^2v_1r +\frac{v_0}{r_0^2}\Big)\cta2\Big\} \, . \nn
\eea
It is timelike in the exterior and becomes null at $r=r _0$. The induced two-metric on a spatial
section of the horizon has a positive definite determinant
\be
\det\, g_{(\theta,\phi)}\big|_{r_0} = \frac{b\bar\omega(r_0)}{g\bar\Xi}\sta2 > 0 \, ,
\ee
For the positive-charge branch, the horizon has a positive area. Its corresponding entropy agrees
with one quarter of the horizon area at $r = r_0$, and is given by
\be\begin{aligned}
S_{\rm BPS}&= \frac{\pi}{g^2\sqrt{B}}\Big[s_1s_2s_3s_4 +(c_1c_2c_3c_4
 -s_1s_2s_3s_4)\frac{s_1s_2s_3s_4}{\sinh D}\sum_{I=1}^4\frac{c_I}{s_I}\Big]^{1/2} \\
&\equiv \frac{4\pi}{g\sqrt{\sum_{I=1}^4\hat{Q}_I}}\sqrt{
\sum_{I=1}^4\frac{\hat{Q}_1\hat{Q}_2\hat{Q}_3\hat{Q}_4}{\hat{Q}_I} -\frac{J}{16g}} \, ,
\end{aligned}\ee
and thermodynamic volume is $V = 4\pi J/(3g)$, where $J$ and $\hat{Q}_I$ take their value in
the BPS limit. This microscopic result extends that obtained from the dual CFT in the pair-wise equal
charge case \cite{JHEP0919079,JKPS76-101} to the more general case with four independent charges.
The conserved quantities for the mass, angular momentum and four electro-magnetic charges resemble
those of the general nonextremal case with the direct replacements: $2m\to\mu$ and $a\to b$.

\section{Separability and hidden symmetry}\label{sec:separability}

The rank-2 Killing-St\"ackel tensor $K^{\mu\nu}$ can be directly read off from the following
expression:
\bea
\Sigma\, K^{\mu\nu}\p_{\mu}\p_{\nu} &=& r^2\Delta_\ta\p_{\ta}^2 +\frac{r^2\Xi}{\sta2}\p_{\phi}^2
 -\frac{r^2\Xi}{g^2\Delta_\ta}\p_t^2
-a^2\cta2\Delta(r)\p_r^2 -\frac{a^2\cta2\, \Xi}{g^2\big[1 +g^2K(r)\big]}\p_t^2 \nn \\
&& -\frac{a^2\cta2\, \Xi}{\Delta(r)\big[1 +g^2K(r)\big]}\big\{\omega(r)\p_t
+a\big[1 +g^2K(r)\big]\p_{\phi}\big\}^2 \, .
\eea
This conformal Killing tensor \cite{CQG27-205009} is responsible for the separability of Hamilton-Jacobi
equation associated with the geodesic motion and the massless Klein-Gordon equation, and hence explains
the hidden symmetry of the spacetime.

\section{Conclusions}\label{sec:conclusion}

In this work, we have presented a new family of nonextremal charged rotating AdS black holes with
four independent charge parameters in four-dimensional $\cN=2$ STU gauged supergravity. It extends
the Cveti\v{c}-Youm \cite{PRD54-2612,NPB717-246} solution that comprises the metric, four Abelian
gauge potentials, three dilatons and three axions to further include a negative cosmological constant.
The metric is given by a very compact form, while the Maxwell vector potentials and their magnetic
duals are regrouped into two sets, and two pairs of dilaton and axion fields are also swapped so
that they can be expressed explicitly in a unified manner.

The solution has neither the Misner string nor the C-metric-like conical defect on the symmetry axes.
It is the only regular black hole in the gauged STU model. Let's present the argument why this is
the case. Further extensions of the present solution would be to add another four charge parameters
and to include the NUT charge parameter, even the acceleration parameter, similar to the general
Pleba\'nski-Demia\'nski family. The resulting solution would be the most general solution yet to be
pursued in the gauged STU model, and has nine solution parameters: the mass, NUT charge, rotation,
acceleration, four pairs of electromagnetic charges, and a nonzero cosmological constant. This generic
solution then contains either the Misner string or the C-metric-like conical defect, or both of them
on the symmetry axes. Even though one can set the total NUT charge to zero so that the Misner string
disappears as did in Ref. \cite{PRD90-025029} by Chow and Comp\`{e}re in the ungauged STU model who
used the Kerr-NUT metric as a starting seed, the obtained solution in the gauged STU model will still
have the conical defect on the symmetry axes since the angular function $\Delta_{\ta}$ will be augmented
by a linear term: $-4g^2Na\cos\ta$, in which $N$ is the total NUT charge.

We have investigated thermodynamic property and hidden symmetry, and derived the BPS condition, which
extends those obtained in the pair-wise equal charge case to the more general case with four independent
charges. This work would attract broad interest from a great deal of different topics, such as shadow,
geodesic motion, hidden symmetry, near-horizon geometry, Kerr/CFT correspondence, holography and
localization, thermodynamic phase transition and topology, etc.

We have investigated thermodynamic property and hidden symmetry, and derived the BPS condition. So
not only is our solution a new background to test the AdS$_4$/CFT$_3$ correspondence and to study the
microscopic entropy of the BPS case, but also our results would invoke a lot of interesting topics,
such as shadow, geodesic motion, hidden symmetry, near-horizon geometry, Kerr/CFT correspondence,
holography localization, thermodynamic phase transition and topology \cite{JHEP0624213,PRD111-L061501,
2607.07364}, etc.

Further generalizations of the present solution had already been finished in 2014. This includes the
AdS extension of the rotating charged Chow-Comp\`{e}re black holes with eight independent charges
in that year, and the extensions of the AdS versions of the Cveti\v{c}-Youm and Chow-Comp\`{e}re
black holes to further include nonzero NUT charge. Due to the complexity of the expressions, we
will present them elsewhere. Still further introducing a nonzero acceleration parameter into these
solutions is under pursued only with partial success. The latter generic solution to four-dimensional
gauged STU supergravity theory should contain thirteen parameters that represent the mass, NUT charge,
rotation and acceleration, four electric and four magnetic charge parameters, and a nonzero cosmological
constant.

\subsection*{Note added}

The general solution was first reported by the second author at the annual
gravitation conference in Hangzhou on June 22, 2015. Recently, L. Ma and H. L\"{u} independently
obtained the solution \cite{MaLu2611}. Both works study its thermodynamics and BPS limit. Ref.
\cite{MaLu2611} also analyzes CTCs, while we provide additional equivalent forms and investigate
hidden symmetries and separability.

\acknowledgments

This work is supported by the National Natural Science Foundation of China (NSFC) under Grants No.
12675077, No. 12375053, No. 12205243 and No. 11275157, by the China Scholarship Council (CSC), by
the Sichuan Science and Technology Program under Grant No. 2026NSFSC0021.


\begin{thebibliography}{99}

\def\CQG{Classical Quantum Gravity\,}
\def\JHEP{J. High Energy Phys.\,}
\def\PRD{Phys. Rev. D\,}
\def\PRL{Phys. Rev. Lett.\,}
\def\NPB{Nucl. Phys. B\,}
\def\PLB{Phys. Lett. B\,}
\def\AP{Ann. Phys. (N.Y.)\,}

\bibitem{LRR23-2}
A. Zaffaroni,
AdS black holes, holography and localization,
\href{https://doi.org/10.1007/s41114-020-00027-8}
{Living Rev. Rel. \textbf{23}, 2 (2020).}

\bibitem{JHEP0516054}
F. Benini, K. Hristov, and A. Zaffaroni,
Black hole microstates in AdS$_4$ from supersymmetric localization,
\href{https://doi.org/10.1007/JHEP05(2016)054}
{\JHEP \textbf{05} (2016) 054}.

\bibitem{PLB771-462}
F. Benini, K. Hristov, and A. Zaffaroni,
Exact microstate counting for dyonic black holes in AdS$_4$,
\href{https://doi.org/10.1016/j.physletb.2017.05.076}
{\PLB \textbf{771}, 462 (2017)}.

\bibitem{JHEP0320081}
J. Nian and L.A. Pando Zayas,
Microscopic entropy of rotating electrically charged AdS$_4$ black holes from field theory localization,
\href{https://doi.org/10.1007/JHEP03(2020)081}
{\JHEP \textbf{03} (2020) 081}.

\bibitem{PRL133-031603}
E. Colombo, S.M. Hosseini, D. Martelli, A. Pittelli, and A. Zaffaroni,
Microstates of accelerating and supersymmetric AdS$_4$ black holes from the spindle index,
\href{https://doi.org/10.1103/PhysRevLett.133.031603}
{\PRL \textbf{133}, 031603 (2024)}.

\bibitem{PRD83-121502}
S.-Q. Wu,
General rotating charged Kaluza-Klein AdS black holes in higher dimensions,
\href{https://doi.org/10.1103/PhysRevD.83.121502}
{\PRD \textbf{83}, 121502(R) (2011)}.

\bibitem{PLB705-383}
S.-Q. Wu,
Two-charged non-extremal rotating black holes in seven-dimensional gauged supergravity:
The Single-rotation case,
\href{https://doi.org/10.1016/j.physletb.2011.10.026}
{\PLB \textbf{705}, 383 (2011)}.

\bibitem{PLB707-286}
S.Q. Wu,
General nonextremal rotating charged AdS black holes in five-dimensional $U(1)^3$ gauged supergravity:
A simple construction method,
\href{https://doi.org/10.1016/j.physletb.2011.12.031}
{\PLB \textbf{707}, 286 (2012).}

\bibitem{JHEP0923143}
N. Bobev, M. David, J. Hong, and R. Mouland,
AdS$_7$ black holes from rotating M5-branes,
\href{https://doi.org/10.1007/JHEP09(2023)143}
{\JHEP \textbf{09} (2023) 143};
\href{https://doi.org/10.1007/JHEP09(2023)198}
{\JHEP \textbf{09} (2023) 198} (erratum).

\bibitem{NPB717-246}
Z.W. Chong, M. Cveti\v{c}, H. L\"{u}, and C.N. Pope,
Charged rotating black holes in four-dimensional gauged and ungauged supergravities,
\href{https://doi.org/10.1016/j.nuclphysb.2005.03.034}
{\NPB \textbf{717}, 246 (2005)}.

\bibitem{PRD89-065003}
D.D.K. Chow and G. Comp\`{e}re,
Dyonic AdS black holes in maximal gauged supergravity,
\href{https://doi.org/10.1103/PhysRevD.89.065003}
{\PRD \textbf{89}, 065003 (2014)}.

\bibitem{FN}
Rotating charged AdS solutions with one parameter less or more than those of Ref. \cite{PRD89-065003}
were successively constructed in \cite{JHEP0114127,JHEP0916088}. It is declared in \cite{JHEP0916088}
that the extra parameter $\beta$ represents a scalar hair, but it is unclear to us whether or not it
is a redundant one.

\bibitem{JHEP0114127}
A. Gnecchi, K. Hristov, D. Klemm, C. Toldo, and O. Vaughan,
Rotating black holes in 4d gauged supergravity,
\href{https://doi.org/10.1007/JHEP01(2014)127}
{\JHEP \textbf{01} (2014) 127}.

\bibitem{JHEP0916088}
S.L. Cacciatori, D. Klemm, and M. Rabbiosi,
Duality invariance in Fayet-Iliopoulos gauged supergravity,
\href{https://doi.org/10.1007/JHEP09(2016)088}
{\JHEP \textbf{09} (2016) 088}.

\bibitem{CQG28-032001}
D.D.K. Chow,
Single-charge rotating black holes in four-dimensional gauged supergravity,
\href{https://doi.org/10.1088/0264-9381/28/3/032001}
{\CQG \textbf{28}, 032001 (2011)}.

\bibitem{CQG28-175004}
D.D.K. Chow,
Two-charge rotating black holes in four-dimensional gauged supergravity,
\href{https://doi.org/10.1088/0264-9381/28/17/175004}
{\CQG \textbf{28}, 175004 (2011)}.

\bibitem{PRL69-1006}
A. Sen,
Rotating Charged Black Hole Solution in Heterotic String Theory,
\href{https://doi.org/10.1103/PhysRevLett.69.1006}
{\PRL \textbf{69}, 1006 (1992)}.

\bibitem{PRL74-1276}
A. Garcia, D. Galtsov, and O. Kechkin,
Class of stationary axisymmetric solutions of the Einstein-Maxwell dilaton-axion field equations,
\href{https://doi.org/10.1103/PhysRevLett.74.1276}
{\PRL \textbf{74}, 1276 (1995).}

\bibitem{PRD103-044014}
D. Wu, S.-Q. Wu, P.X. Wu, and H.W. Yu,
Aspects of the dyonic Kerr-Sen-AdS$_4$ black hole and its ultraspinning version,
\href{https://doi.org/10.1103/PhysRevD.103.044014}
{\PRD \textbf{103}, 044014 (2021)}.

\bibitem{PR316-1}
D. Youm,
Black holes and solitons in string theory,
\href{https://doi.org/10.1016/S0370-1573(99)00037-X}
{Phys. Rep. \textbf{316}, 1 (1999).}

\bibitem{CQG31-022001}
D.D.K. Chow and G. Comp\`{e}re,
Seed for general rotating non-extremal black holes of N=8 supergravity,
\href{https://doi.org/10.1088/0264-9381/31/2/022001}
{\CQG \textbf{31}, 022001 (2014)}.

\bibitem{PRD90-025029}
D.D.K. Chow and G. Comp\`{e}re,
Black holes in N=8 supergravity from SO(4,4) hidden symmetries,
\href{https://doi.org/10.1103/PhysRevD.90.025029}
{\PRD \textbf{90}, 025029 (2014)}.

\bibitem{PRD54-2612}
M. Cveti\v{c} and D. Youm,
Entropy of non-extreme charged rotating black holes in string theory,
\href{https://doi.org/10.1103/PhysRevD.54.2612}
{\PRD \textbf{54}, 2612 (1996).}

\bibitem{JGP23-111}
L. Andrianopoli, M. Bertolini, A. Ceresole, R. D'Auria, S. Ferrara, P. Fr\'{e}, and T. Magri,
N=2 supergravity and N=2 superYang-Mills theory on general scalar manifolds: Symplectic
covariance, gaugings and the momentum map,
\href{https://doi.org/10.1016/S0393-0440(97)00002-8}
{J. Geom. Phys. \textbf{23}, 111 (1997)}.

\bibitem{NPB558-96}
M. Cveti\v{c}, M.J. Duff, P. Hoxha, J.T. Liu, H. L\"{u}, J.X. Lu,
R. Martinez-Acosta, C.N. Pope, H. Sati, and T.A. Tran,
Embedding AdS black holes in ten-dimensions and eleven-dimensions,
\href{https://doi.org/10.1016/S0550-3213(99)00419-8}
{\NPB \textbf{558}, 96 (1999)}.

\bibitem{NPB554-237}
M.J. Duff and J.T. Liu,
Anti-de Sitter black holes in gauged N=8 supergravity,
\href{https://doi.org/10.1016/S0550-3213(99)00299-0}
{\NPB \textbf{554}, 237 (1999)}.

\bibitem{hep-th/0504080}
M. Cveti\v{c}, G.W. Gibbons, H. L\"u, and C.N. Pope,
Rotating black holes in gauged supergravities: Thermodynamics, supersymmetric limits,
topological solitons and time machines,
\href{https://arXiv.org/abs/hep-th/0504080}
{hep-th/0504080}.

\bibitem{CQG17-399}
M.M. Caldarelli, G. Cognola, and D. Klemm,
Thermodynamics of Kerr-Newman-AdS black holes and conformal field theories,
\href{https://doi.org/10.1088/0264-9381/17/2/310}
{\CQG \textbf{17}, 399 (2000)}.

\bibitem{PRD73-104036}
W. Chen, H. L\"{u}, and C. N. Pope,
Mass of rotating black holes in gauged supergravities,
\href{https://doi.org/10.1103/PhysRevD.73.104036}
{\PRD \textbf{73}, 104036 (2006)}.

\bibitem{JHEP0919079}
D. Cassani and L. Papini,
The BPS limit of rotating AdS black hole thermodynamics,
\href{https://doi.org/10.1007/JHEP09(2019)079}
{\JHEP \textbf{09} (2019) 079}.

\bibitem{JKPS76-101}
S. Choi, C. Hwang, S. Kim, and J. Nahmgoong,
Entropy Functions of BPS Black Holes in AdS$_4$ and AdS$_6$,
\href{https://doi.org/10.3938/jkps.76.101}
{J. Korean Phys. Soc. \textbf{76}, 101 (2020)}.

\bibitem{CQG27-205009}
D.D.K. Chow,
Symmetries of supergravity black holes,
\href{https://doi.org/10.1088/0264-9381/27/20/205009}
{\CQG \textbf{27}, 205009 (2010).}

\bibitem{JHEP0624213}
D. Wu, S.-Y. Gu, X.-D. Zhu, Q.-Q. Jiang, and S.-Z. Yang,
Topological classes of thermodynamics of the static multi-charge AdS black holes in gauged supergravities: Novel temperature-dependent thermodynamic topological phase transition,
\href{https://doi.org/10.1007/JHEP06(2024)213}
{\JHEP \textbf{06} (2024) 213}.

\bibitem{PRD111-L061501}
D. Wu, W. Liu, S.-Q. Wu, and R. B. Mann,
Novel topological classes in black hole thermodynamics,
\href{https://doi.org/10.1103/PhysRevD.111.L061501}
{\PRD \textbf{111}, L061501 (2025)}.

\bibitem{2607.07364}
D. Wu,
Projections and topological classification of black hole thermodynamics,
\href{https://arxiv.org/abs/2607.07364}{arXiv:2607.07364}.

\bibitem{MaLu2611}
L. Ma and H. L\"{u},
Four-charge Rotating AdS Black Holes in Gauged STU Supergravity,
\href{https://arxiv.org/abs/2611.13074}{arXiv:2611.13074}.

\end{thebibliography}
\end{document}